# Electrical conductivity of MgSiO$_3$ at high temperatures and pressures: implications for the Earth's mantle

Samuel S. M. Santos,[1] Lucy V. C. Assali,[2] and João F. Justo[1]

[1]*Escola Politécnica, Universidade de São Paulo,*
*CEP 05508-010, São Paulo, SP, Brazil*

[2]*Instituto de Física, Universidade de São Paulo,*
*CEP 05315-970, São Paulo, SP, Brazil*

(Dated: February 5, 2024)

The electrical conductivity of magnesium silicate MgSiO$_3$ has been studied, using the framework of the first-principles density functional theory and the Boltzmann transport theory, under the thermodynamic conditions of the Earth's lower mantle. We find that the conductivity of pristine MgSiO$_3$ depends strongly on the structural phase of the material, as well as on temperature and pressure. The conductivity of the perovskite phase increases with increasing pressure (depth of the lower mantle) up to 90 GPa, then decreases at higher pressures due to a change in the material's band gap transition from direct to indirect. Finally, the structural phase transition that MgSiO$_3$ undergoes near the bottom of the lower mantle, from perovskite to post-perovskite, causes an increase in the conductivity of MgSiO$_3$, which should contribute to the increase in the electrical conductivity of the Earth's mantle under the thermodynamic conditions of the Earth's D″ layer.

# I. INTRODUCTION

Studying the electrical conductivity of minerals, as well as other physical properties, can provide important information about the Earth's interior. Sounding methods have estimated that the electrical conductivity increases from about $10^{-4}$- $10^{-2}$ to $10^0$ S/m from the crust to the top of the lower mantle[1]. Nevertheless, it is estimated that the conductivity does not vary significantly in the lower mantle, although geophysical modeling has proposed the possible existence of a highly conductive layer at its base. Furthermore, the extreme conditions of high pressures and high temperatures affect the conductivity of lower mantle minerals. Additionally, conductivity depends on the chemical composition, formation of defects and impurities, iron spin transition, oxygen fugacity, and water content that may occur in these minerals[1–3].

Electrical conductivity measurements performed in laboratory can provide valuable information about mantle structure and composition. In general, the various results from different laboratories have indicated a contouring trend with respect to the electrical conductivity of mantle minerals. Most of the mineral phases that make up the Earth's mantle exhibit semiconductor behavior[3]. The mechanisms that control the electrical conductivity of mantle minerals are associated with electron, proton and small polaron conduction, while ion mobility may also contribute at high temperatures near the mineral's melting point[1,3]. To distinguish the various charge transport mechanisms in mantle minerals, it is necessary to measure their electrical conductivity over a wide temperature range.

The structural phase transition that mantle minerals undergo has a significant effect on the Earth's electrical conductivity distribution. Laboratory measurements have shown that olivine can have high conductivity values in the oxidized upper and shallow mantle[3]. With increasing depth, the phase transition from olivine to wadsleyite and from wadsleyite to ringwoodite along the adiabatic geotherm causes the conductivity to increase[1]. The conductivity may further increase with the transition to the post-spinel phase. Finally, the conductivity may vary due to the iron spin transition in the middle of the lower mantle and then increase substantially in the D″ layer.

Silicates are the major phase in the Earth's lower mantle, these minerals can store iron, aluminum, and carbon in the interior of our planet[1,4–8]. In general, silicates are not good conductors at low temperatures, but at high temperatures they present a higher electrical conductivity[9]. They act as insulators at room temperature, however behaving as semiconductors under mantle conditions[1–3]. $MgSiO_3$ is believed to be the most abundant mineral in the Earth's lower mantle and its electrical conductivity has been extensively studied by measurements in several laboratories[10–14]. These measurements have been performed considering impurities, such as iron, aluminum, and sodium, which increase the conductivity of $MgSiO_3$. It is already well established, from both experimental and theoretical viewpoints, that in the lower mantle the pristine $MgSiO_3$ undergoes a phase transition, from perovskite to post-perovskite[6,15–21], and that this transition is directly associated with the D″ discontinuity of the mantle. Taking into account these various works, both experimental and theoretical, the pressure at which this phase transition occurs is within the 98-125 GPa range.

From a modeling point of view, the electrical conductivity of several minerals has been obtained using the Boltzmann transport theory, based on first-principles calculations. In general, the effect of temperature causes conductivity to increase in perovskite structures. Nonetheless, the effect of increasing pressure makes the electrical conductivity decrease for the $BaSiO_3$ perovskite[22] and increase for the $BaGeO_3$ perovskite[23]. The band gap width of these two perovskites increases with increasing pressure and, in addition, their nature change, undergoing from an indirect transition $(R-\Gamma)$ to a direct one $(\Gamma-\Gamma)$. Furthermore, by analyzing the effect of chemical substitution in the $MgSiO_3$, $MgGeO_3$, and $MgSnO_3$ cubic perovskite series[24], it can be observed that the conductivity is directly related to the band gap width value of these minerals, i.e., $MgSiO_3$ has wider band gap and higher conductivity than $MgGeO_3$ and $MgSnO_3$, at 0 GPa.

This work is focused on the electrical conductivity of pristine $MgSiO_3$. The study was carried out considering the effects of high pressure and high temperature in order to take into account the conditions of the lower mantle on the electronic conductivity of $MgSiO_3$, associated with electronic transport, as well as the effect of the perovskite to post-perovskite phase transition. Furthermore, the influence of the electronic band structure characteristics on the electrical conductivity of $MgSiO_3$ is also discussed.

## II. METHODOLOGY AND COMPUTATIONAL DETAILS

The calculations were performed using the Quantum ESPRESSO (QE) computational package[25]. The electronic interactions were described within the density functional theory (DFT), considering the exchange-correlation (XC) functional based on the generalized gradient approximation (GGA) specified by Perdew–Burke–Ernzerhof (PBE)[26]. The electronic wave functions were expanded using the projected augmented wave (PAW) method[27], with a plane-wave cutoff of 1200 eV. The valence electronic configurations were described by $3s^2 3p^2$ for silicon (Si), $2s^2 2p^6 3s^2 3p^0$ for magnesium (Mg), and $2s^2 2p^4$ for oxygen (O) atoms[28]. Some tests were performed with the Tran-Blaha modified Becke and Johnson (TB-mBJ) XC-potential approximation, which was implemented together with the GGA-PBE, as it is well established in the literature that this TB-mBJ functional gives a better description of the band gap width values in semiconducting materials. Since the electronic properties are strongly dependent on the material's band structure features, it is important to understand the changes that are introduced in the properties obtained with the GGA-PBE approach when the TB-mBJ scheme is used.

$MgSiO_3$ was investigated in perovskite and post-perovskite structures. The perovskite structure was considered in cubic (PV-C) and orthorhombic (PV-O) phases, which belong to the $Pm\bar{3}m$ (#221) and $Pbnm$ (#62) space groups, respectively. The post-perovskite structure was considered in the orthorhombic phase (PPV-O), whose space group is $Cmcm$ (#63). The PV-C phase was only simulated at zero pressure, as this phase exists only at very low pressures, and it was used as the reference model at 0 GPa for the other two structures studied here, the PV-O and the PPV-O. Beyond the low-pressure regime, the PV-O and PPV-O phases were simulated from 0 to 150 GPa. These two phases play a very important role in this pressure regime and are fundamental for understanding the Earth's mantle structure and composition. For each pressure, the structures were optimized by using the damped variable cell shape molecular dynamics method[29]. The atomic positions were considered converged when all forces acting on atoms were smaller than 0.01 eVÅ$^{-1}$. The Brillouin zones for computing the electronic properties were sampled by a $(8 \times 8 \times 8)$ $k$-mesh for PV-C, $(4 \times 4 \times 4)$ for PV-O and $(6 \times 6 \times 6)$ for PPV-O.

The electrical conductivity, as a function of the temperature, was obtained using the Boltzmann transport theory based on first-principles calculations, as implemented in the

BoltzTraP computational code[30]. This scheme uses a smoothed Fourier interpolation of the bands to obtain the semi-classical transport coefficients as a function of the relaxation time $\tau$, the electronic-scattering rate. It uses the constant relaxation time approximation (CRTA) as the default solution, which treats $\tau$ as a constant since it cannot be derived from the band structure analysis. It is worth mentioning that to obtain the relaxation time as a function of temperature, it is necessary to obtain, for each pressure, the phonon spectrum, which demands considerable additional computational resources, which is not the scope of this work.

In this work, $MgSiO_3$ was studied in its pristine structure, and the electronic contribution to conductivity was computed. Other important contributions to the conductivity of a mineral under mantle temperature and pressure conditions were not taken into account, including defects and impurities, such as hydrogen, oxygen fugacity, or ferric iron, among others[1]. For most thermodynamic conditions of the lower mantle, the electronic contribution is the prevailing contribution, which is described by the methodology of this investigation, although for the extreme conditions of the bottom of the mantle, the other contributions start to play some role in conductivity.

The non-self-consistent calculations, which must be performed after the simulations have been converged and all theoretical structural parameters have been obtained, provide the input data for the BoltzTraP code, and were performed with the following *k*-mesh sampling: (20×20×20) for PV-C, (12×12×12) for PV-O and (16×16×16) for PPV-O. The temperature range from 50 to 3500 K, in 50 K steps, was used to obtain the electrical conductivity, and for each structural phase and pressure, it was obtained at the corresponding Fermi level.

## III. RESULTS

Figure 1 (a) displays the electronic band structures of MgSiO$_3$ in the PV-C, PV-O, and PPV-O phases at the ground state, i.e, at $T = 0$K and null pressure. Although in the low-pressure regime, MgSiO$_3$ exhibits the PV-O phase, the results of the hypothetical PV-C and PPV-C phases at 0 GPa, were also simulated in order to provide a reference for comparison. The PV-C and PV-O perovskite phases exhibit an indirect transition band gap, while the PPV-O phase exhibits a direct one. Detailed information on the band gap widths and band structures are given in Table I.

The projected density of states (PDOS) of these phases shows that the valence bands are mostly derived from the O $p$-orbitals, while the bottom of the conduction bands has predominantly Mg $s$-type characteristics. Furthermore, both Si and O $p$-orbitals contribute to the conduction band composition. The results of the MgSiO$_3$ in the PV-C phase are in excellent agreement with previous DFT results[24].

Figure 1 (b) shows the electrical conductivity as a function of temperature for the three phases of MgSiO$_3$. Near the ground state, 50 K and 0 GPa, the PV-C phase has higher electrical conductivity than the PV-O and PPV-O phases. The increase in the electrical conductivities as a function of these phases follows the inverse relationship of the band gap widths. The PV-C phase has a narrower band gap and a higher conductivity than the PPV-O one, while the PPV-O phase has a narrower band gap and a higher conductivity than the PV-O one, as listed in Table I.

According to the results in Figure 1 and Table I, the gap width directly affects the MgSiO$_3$ conductivity, which is also affected by the curvature of the VBM (valence band maximum) and whether the band gap transition is direct or indirect. In the PPV-O phase, the conductivity is significantly attenuated because its VBM is not as sharp as that of the PV-C and PV-O phases. At 300 K and at the Fermi level, the electrical conductivity of MgSiO$_3$ in the PV-C phase is $1.99 \times 10^{19}$ S/ms, which is in excellent agreement with previous results obtained by DFT calculations[24].

As previously mentioned, the thermoelectric properties of MgSiO$_3$, as well as its electrical conductivity, depend strongly on the electronic band structure composition. In the lower mantle, this mineral presents two phases that have orthorhombic structures: perovskite and post-perovskite. Figure 2 shows the changes in the VBM and CBM (conduction band minimum) regions of the band structures under high pressures, 90 and 120 GPa, of these two phases: (a) PV-O and (b) PPV-O. At both pressures, the effect of increasing pressure causes the band gap width to increase. Moreover, in the perovskite phase, the gap passes from an indirect to a direct transition: at 90 GPa the gap direction is $\sim$ Z $-$ $\Gamma$, while at 120 GPa the transition becomes $\Gamma - \Gamma$. In the post-perovskite phase, the gap transition is always indirect: $\Gamma - $Y. It can also be noted that the increase in gap width under increasing pressure is greater in the perovskite phase than in the post-perovskite one. However, the VBM curvature of the PV-O phase becomes higher at higher pressures, while in the PPV-O phase, the curvature of the bands is almost unchanged.

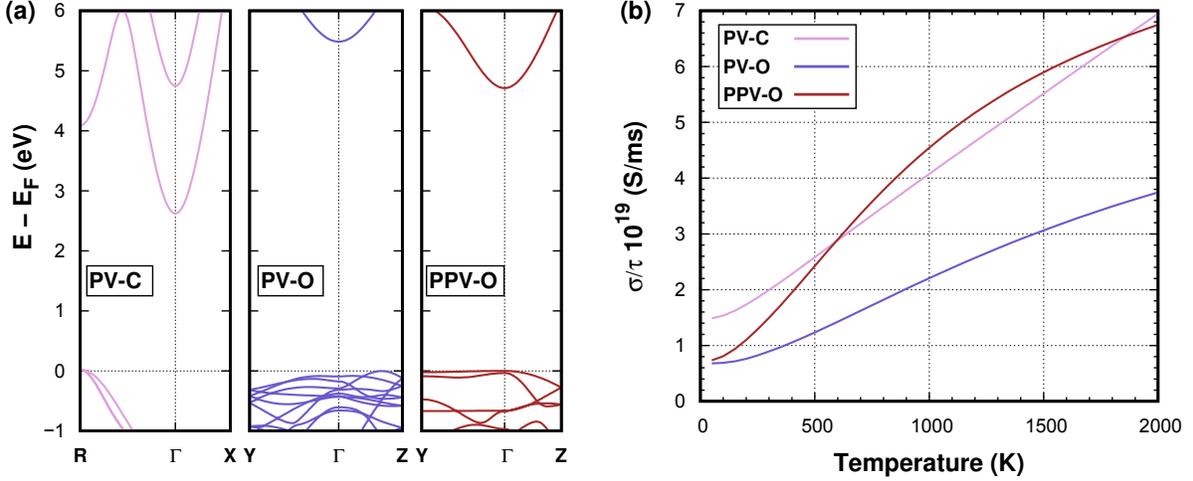

Figure 1. Properties of MgSiO$_3$ in the PV-C, PV-O, and PPV-O structural phases, at null pressure, for (a) ground state electronic band structures at $T = 0$ K and (b) electrical conductivity as a function of temperature.

Table I. Ground state band gap (E$_g$), gap transition direction, and electrical conductivity of MgSiO$_3$ in the PV-C, PV-O, and PPV-O structural phases at 50 K and at the Fermi level.

| Phases | E$_g$ (eV) | gap transition | $\sigma/\tau$ ($10^{19}$ S/ms) |
|---|---|---|---|
| PV-C | 2.60 | R $-$ $\Gamma$ | 1.49 |
| PV-O | 5.50 | $\sim$ Z $-$ $\Gamma$ | 0.68 |
| PPV-O | 4.71 | $\Gamma$ $-$ $\Gamma$ | 0.73 |

Figure 3 (a) shows the electrical conductivity of PV-O and PPV-O phases of the MgSiO$_3$ as a function of pressure, at the Fermi level and at 50, 300, and 500 K. Pressure affects the conductivity of the PV-O and PPV-O phases in different ways. First, for the PV-O phase

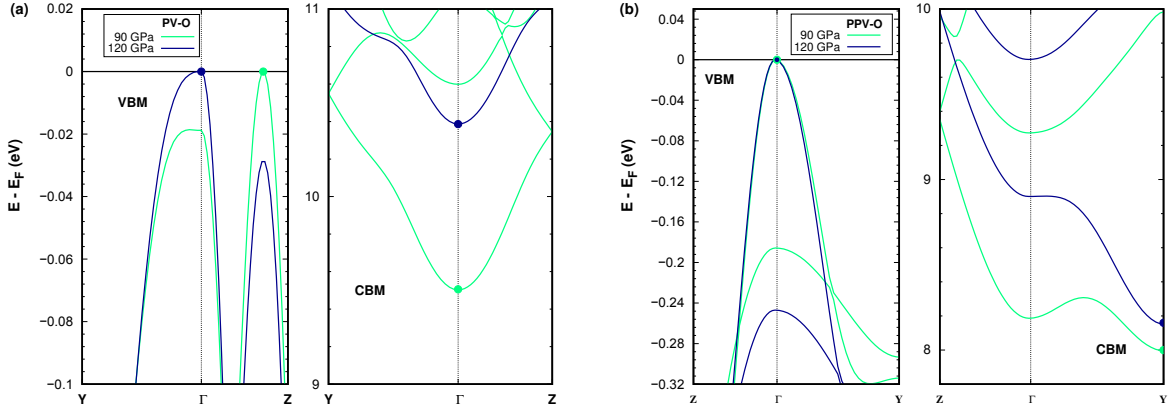

Figure 2. Electronic band structures, under pressure, of the MgSiO$_3$ on the (a) PV-O phase and (b) PPV-O phase, at 90 GPa and 120 GPa. The valence band maximum is highlighted, for each pressure, on the left side, while on the right side, the conduction band minimum is highlighted.

and between 0 and 90 GPa, the conductivity increases with increasing pressure.

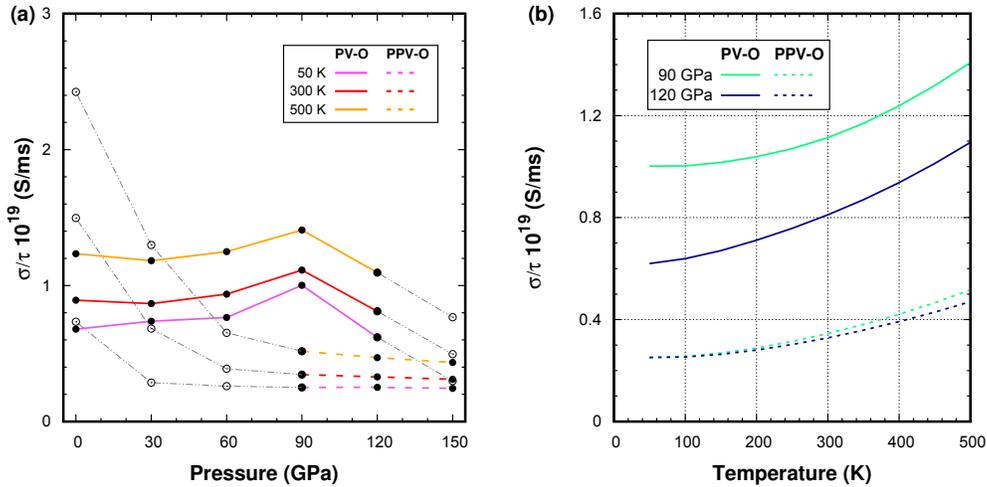

Figure 3. (a) MgSiO$_3$ electrical conductivity in the PV-O and PPV-O phases as a function of pressure at 50, 300, and 500 K. (b) Change in electrical conductivity of MgSiO$_3$ in the PV-O and PPV-O phases as a function of temperature at 90 GPa and 120 GPa.

Then, between 90 and 150 GPa, the conductivity decreases. In other words, at 90 GPa the slope of the conductivity versus pressure curve changes sign, from positive to negative. The main reason for this modification in the conductivity trend is related to the gap transition nature change that the PV-O phase undergoes between 90 and 120 GPa. It can also be observed that for the PV-O phase, for any pressure, the effect of increasing temperature causes the conductivity to increase. On the other hand, for the PPV-O phase, the conductivity decreases or remains unchanged with increasing pressure, for any temperature.

For this phase, the gap width remains practically constant under the pressure conditions in which this phase exists, between 90 and 150 GPa, and also no gap transition nature change is observed. Furthermore, the curvature of the VBM of the PPV-O phase does not change significantly. Therefore, the pressure does not cause relevant variations in the electrical conductivity of $MgSiO_3$ in this phase because pressure also does not meaningfully change the characteristics of the electronic band structure of the PPV-O phase, as it occurs in the PV-O.

Figure 3 (b) shows the $MgSiO_3$ electrical conductivity, as a function of temperature, in the PV-O and PPV-O phases, at the Fermi level and 90 and 120 GPa, for temperatures up to 500 K. The effect of increasing pressure, for both phases, causes a decrease in conductivity. For the PV-O phase, at 50 K, electric conductivity decreases from $1.0 \times 10^{19}$ S/ms, at 90 GPa, to $0.6 \times 10^{19}$ S/ms, at 120 GPa. The difference in the conductivity between these two pressures remains practically the same for the entire temperature range studied, being $0.4 \times 10^{19}$ S/ms at 50 K and $0.3 \times 10^{19}$ S/ms at 500 K. On the other hand, different behavior is observed for the PPV-O phase, that at low temperatures, the conductivity does not change with increasing pressure, but varies in the high-temperature regime, being 0 at 50 K and $0.05 \times 10^{19}$ S/ms at 500 K. It is important to point out that the conductivity of the PV-O phase has a greater variation with pressure than that of the PPV-O phase, which remains almost constant up to $\sim$ 500 K. This behavior is due to the major change in the width and nature of the band gap that the PV-O phase experiences due to the pressure effects.

Figure 4 (a) depicts the $MgSiO_3$ conductivity as a function of pressure of the PV-O and PPV-O phases, at high temperatures. In this temperature regime, the conductivity of the PV-O phase varies more linearly with pressure than in the low-temperature regime. Despite that, the sign change of the curve slope still occurs at 90 GPa, as well as at low temperatures regime, as shown in Figure 3 (a). This behavior suggests that, even at high temperatures, the conductivity of the PV-O phase depends on the width and kind of the band gap transition. In turn, the conductivity of the PPV-O phase varies greatly with pressure variation at high temperatures, contrary to what was observed in the low-temperature regime. It can be observed that from 90 GPa, up to 150 GPa, both PV-O and PPV-O phases show a negative slope for the conductivity versus pressure. Moreover, those curves intersect at high temperatures, an effect that does not occur at low temperatures, as can be observed in Figure 3. The intersection of the conductivity versus pressure curves of the two phases of $MgSiO_3$ occurs close to the thermodynamic conditions of the Earth's D″ layer.

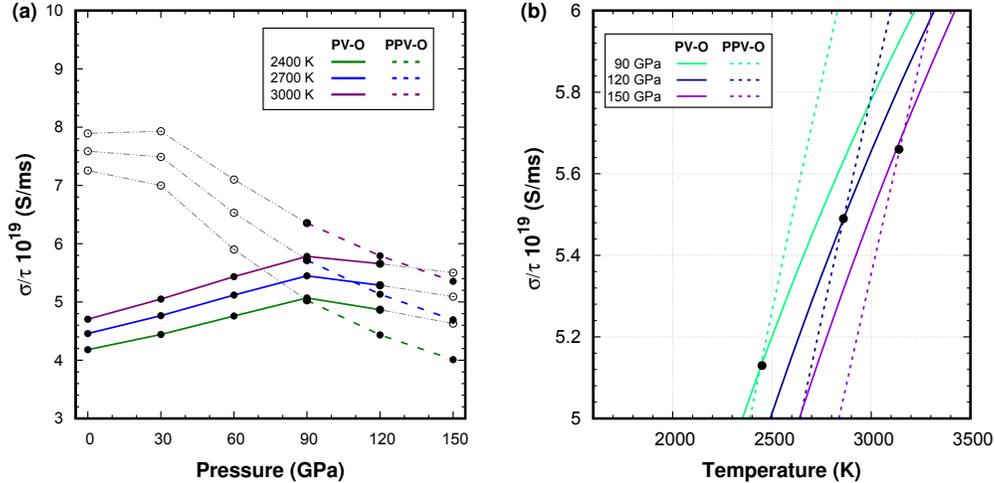

Figure 4. (a) Electrical conductivity versus pressure of $MgSiO_3$ in the PV-O and PPV-O phases at several temperatures. (b) Electrical conductivity versus temperature of $MgSiO_3$ in the PV-O and PPV-O phases at 90, 120, and 150 GPa. The intersections of the PV-O and PPV-O phase conductivities are highlighted by solid black circles.

Figure 4 (b) displays the electrical conductivity as a function of temperature of the PV-O and PPV-O phases, at the Fermi level and at 90, 120, and 150 GPa, at high temperatures. The intersections of the conductivity versus temperature curves of the PV-O and PPV-O phases are highlighted by solid black circles. Only at 2450 K and 90 GPa that the PPV-O phase starts to have higher conductivity than the PV-O one. Furthermore, if the pressure is 120 GPa, then the temperature at which this change takes place is 2860 K. Around 120 GPa and 125 GPa, experimental data have shown that the structural phase transition from PV-O to PPV-O occurs at 2000 K and 2500 K, respectively[21]. Therefore, the temperatures at which the structural phase transition occurs are lower than those at which the conductivities of the PV-O and PPV-O phases cross, for a given pressure. That is, our results show an overestimation of the phase transition from PV-O to PPV-O in terms of electrical conductivity. However, as mentioned above, it is recognized that the GGA-PBE exchange-correlation potential approach overestimates the structural phase transition pressure values and, therefore, it is expected that the results obtained in this work overestimate the electrical conductivity crossing points for these two phases of $MgSiO_3$.

Since the electrical conductivity is strongly dependent on the band structure of the mineral, some simulations were carried out using the TB-mBJ exchange-correlation potential, at null pressure, which is already well known to improve the electronic band gap width. Our results using this TB-mBJ approximation show that the conductivity of the PV-O phase is overestimated by 0.8, while the PPV-O phase is overestimated by 0.5, in units of $10^{19}$ S/ms. In other words, at 2400 K, for example, the conductivity of the PV-O phase changes from $4.2 \times 10^{19}$ S/ms (Figure 4 (a)) to $3.4 \times 10^{19}$ S/ms, while the conductivity of the PPV-O phase changes from $7.2 \times 10^{19}$ S/ms (Figure 4 (a)) to $6.7 \times 10^{19}$ S/ms. This correction means that the temperature at which the conductivities of these two phases intersect is lower than the corresponding ones displayed in Figure 4 (b), i.e., the intersections at 90 and 120 GPa would be 2150 K and 2500 K, respectively, by using TB-mBJ potential. Therefore, the correct treatment of the electronic structure of the $MgSiO_3$ mineral indicates that the conductivity crossing of both PV-O and PPV-O phases occurs close to the pressure and temperature that this mineral undergoes a structural phase transition.

In the region between 90 and 150 GPa, increasing temperature causes the conductivity of the PPV-O phase to become greater than that of the PV-O one. Therefore, the phase transition that $MgSiO_3$ undergoes in the Earth's mantle causes the conductivity of this region to increase with increasing pressure or depth. However, such a comparison should be taken with caution, since several phenomena occur simultaneously in the 60 GPa region of the mantle, such as the high-to-low iron spin transition, which is incorporated in both MgO and $MgSiO_3$. In general, the $MgSiO_3$ conductivity increases up to a certain mantle depth, corresponding to 90 or 120 GPa, then the conductivity undergoes a positive shift due to the PV-O to PPV-O phase transition, after which it decreases more smoothly to the bottom of the Earth's lower mantle.

Table II presents a collection of experimental data on the silicate perovskite $MgSiO_3$ electrical conductivity under several pressure and temperature conditions, which can store iron, aluminum, and sodium[4,5,11]. The amount of iron in perovskite silicates ranges from 0 to 10%[10–14] and 2.8 wt.% of aluminum[12]. Hereupon, we use the experimental conductivity of silicate perovskites containing impurities to estimate the relaxation time value of pure $MgSiO_3$. For each pressure value and the average value of the experimental temperature range reported in Table II, an estimate of the relaxation time $\tau_{est}$ is obtained with the following expression: $\tau_{est} = \sigma_{exp}/(\sigma/\tau)$. By using this expression, we obtain $\tau_{est} \approx 10^{-17}$ s. Bearing in mind this $\tau_{est}$ as the relaxation time value of $MgSiO_3$ for both phases, PV-O and PPV-O, the increase in the conductivity with mantle depth would be 99 S/m at 2700 K and under pressures between 0 and 90 GPa. Now, considering the structural phase transition effect, at 2700 K and 90 GPa, the conductivity increase is 27 S/m. Therefore, taking into account both effects, the increase in the conductivity of the $MgSiO_3$ under these pressure and temperature conditions is 126 S/m. This value is in line with the increase in the Earth's mantle conductivity from top to bottom, which is estimated to be from $10^2$ to $10^4$ S/m[1].

Table II. Experimental data of silicate perovskite $MgSiO_3$ electrical conductivity $\sigma$ (S/m) under several pressure and temperature conditions, that may contain certain amounts of iron, aluminum, or sodium.

| $\Delta P$ (GPa) | $\Delta T$ (K) | $\ln(\sigma)$ | Reference |
|---|---|---|---|
| 23 | 400 - 900 | 0.60 | Katsura *et al.* (1998)[10] |
| 23 | 1500 - 2000 | 2.30 | Katsura *et al.* (1998)[10] |
| 24 | 1300 - 1800 | 3.92 (3) | Dobson (2003)[11] |
| 24 | 1800 - 2200 | 9.40 (40) | Dobson (2003)[11] |
| 25 | 1673 - 1873 | 1.12 (12) | Xu *et al.* (1998b)[12] |
| 25 | 1673 - 1873 | 1.87 (11) | Xu *et al.* (1998b)[12] |
| 25 | 350 - 1100 | 0.91 (2) | Yoshino *et al.* (2008c)[13] |
| 25 | 1750 - 2000 | 2.64 (6) | Yoshino *et al.* (2008c)[13] |
| 1.2 - 40 | 293 - 673 | 1.92 (68) | Shankland *et al.* (1993)[14] |

The conductivity values could also be estimated through the band gap width variation of the $MgSiO_3$. If only the effect of the gap width variation is considered, the increase in the conductivity ($\Delta \ln \sigma$), which is produced by the band gap width reduction ($\Delta E_g$) due to the structural transition from PV-O to PPV-O phases, can be estimated through the relation: $\Delta \ln \sigma = -\Delta E_g/(2k_B T)$, where $k_B$ is the Boltzmann constant. The static calculations of the $MgSiO_3$ perfect crystal show that the band gap widths of the PPV-O phase, at pressures between 90 and 150 GPa, are 16% to 25%, respectively, narrower than that of the PV-O phase, and at 2500 K, these gap width reductions increase the conductivity $\Delta \ln \sigma$ from 3.5 to 6.4. These results agree with the increase of $\Delta \ln \sigma = 6.68$ obtained by another theoretical calculation at 125 GPa and 2500 K[31] and indicate that the high conductivity of the $MgSiO_3$ in the PPV-O phase is a result of the band gap reduction due to the transition from PV-O to PPV-O phase. An increase of $\Delta \ln \sigma = 3.5$ means an increase of $\Delta \sigma = 30$ S/m, which agrees with the value of 27 S/m obtained previously by estimating $\tau_{est}$.

## IV. CONCLUSIONS

Here, the electrical conductivity of pristine $MgSiO_3$, obtained using Boltzmann transport theory based on DFT calculations, is shown to be greatly affected by pressure and temperature conditions. The simulations took into account the extreme conditions to which this mineral is subjected in the Earth's mantle. Under these conditions, $MgSiO_3$ undergoes a structural phase transition, which plays an important role in the conductivity behavior of this mineral.

The electronic band structure of $MgSiO_3$ in the PV-O phase is strongly affected by increasing pressure and, consequently, by the increase in depth of Earth's lower mantle, modifying the width and transition direction of its band gap. Thus, as expected, the electrical conductivity of the PV-O phase is also greatly affected by the increase in pressure, since the electronic bands play a key role in the electronic conductivity of a semiconducting mineral. On the other hand, as the electronic bands of the PPV-O phase are not significantly affected by pressure, their conductivity remains practically constant at the thermodynamic conditions of the lower mantle.

As the structures of the electronic bands were obtained for the ground state, for each pressure, the above conclusions refer to the low-temperature limit, since the effect of temperature on the electronic band has not been taken into account. In the high-temperature limit, although the electronic band should not affect the conductivity much, the change in the material's band gap transition from direct to indirect, after 90 GPa, still contributes to a decrease in the conductivity of the PV-O phase. This favors the conductivity of the PPV-O phase, which becomes greater than that of the PV-O phase. Therefore, it is reasonable to assume that the structural transition from the perovskite to post-perovskite phase of $MgSiO_3$ contributes to the increase in the electrical conductivity of the Earth's lower mantle.

Finally, it is estimated that the electrical conductivity increases with depth and it is of the order of $10^0$ S/m in the lower mantle. This is consistent with the increase in conductivity of pristine $MgSiO_3$, with increasing pressure and temperature, as well as with the structural phase transition. However, a more accurate calculation needs to take into account other electrical conductivity mechanisms. Considering the additive nature of the distinct electrical conductivity mechanisms, our results on the electronic contribution represent a lower limit of this conductivity. Other contributions to the electrical conductivity of these min-erals are expected to play an important role at high temperatures and pressures. Further improvements in the description of conductivity will require moving beyond the constant relaxation time approximation, which was used in this investigation.

# V. REFERENCES

[1] T. Yoshino, Surv. Geophys. **31**, 163 (2010).

[2] K. Shun-Ichiro, Earth Planet. Sci. Lett. **301**, 413 (2011).

[3] X. Yang, H. Liu, and X. Zhao, Contrib. Mineral. Petrol **178**, 18 (2023).

[4] E. Ito and E. Takahashi, J. Geophys. Res. Solid Earth **94**, 10637 (1989).

[5] B. Wood and D. Rubie, Science **273**, 1522 (1996).

[6] S. S. M. Santos, M. L. Marcondes, J. F. Justo, and L. V. C. Assali, Earth Planet. Sci. Lett. **506**, 1 (2019).

[7] F. Gaillard, M. Malki, G. Iacono-Marziano, M. Pichavant, and B. Scaillet, Science **322**, 1363 (2008).

[8] S. S. M. Santos, M. L. Marcondes, J. F. Justo, and L. V. C. Assali, Phys. Earth Planet. Inter. **299**, 106327 (2020).

[9] K. Han and S. M. Clark, Solid Earth Sci. **6**, 111 (2021).

[10] T. Katsura, K. Sato, and E. Ito, Nature **395**, 493 (1998).

[11] D. Dobson, Phys. Earth Planet. Inter. **139**, 55 (2003).

[12] Y. Xu, C. McCammon, and B. T. Poe, Science **282**, 922 (1998).

[13] T. Yoshino, D. Yamazaki, E. Ito, and T. Katsura, Geophys. Res. Lett. **35**, L22303 (2008).

[14] T. J. Shankland, J. Peyronneau, and J.-P. Poirier, Nature **366**, 453 (1993).

[15] A. R. Oganov and S. Ono, Nature **430**, 445 (2004).

[16] J. Tsuchiya, T. Tsuchiya, and R. M. Wentzcovitch, Phys. Rev. B **72**, 020103 (2005).

[17] T. Tsuchiya, J. Tsuchiya, K. Umemoto, and R. M. Wentzcovitch, Earth Planet. Sci. Lett. **224**, 241 (2004).

[18] N. Guignot, D. Andrault, G. Morard, N. Bolfan-Casanova, and M. Mezouar, Earth Planet. Sci. Lett. **256**, 162 (2007).

[19] Z.-J. Liu, X.-W. Sun, C.-R. Zhang, J.-B. Hu, L.-C. Cai, and Q.-F. Chen, Bull. Mater. Sci. **35**, 665 (2012).

[20] G. Fiquet, A. Dewaele, D. Andrault, M. Kunz, and T. Le Bihan, Geophys. Res. Lett. **27**, 21 (2000).

[21] M. Murakami, K. Hirose, K. Kawamura, N. Sata, and Y. Ohishi, Science **304**, 855 (2004).

[22] Q. Mahmood, M. Hassan, S. Ahmad, A. Shahid, and A. Laref, J. Phys. Chem. Solids **120**, 87 (2018).

[23] N. Noor, Q. Mahmood, M. Hassan, A. Laref, and M. Rashid, J. Mol. Graph. Model. **84**, 152 (2018).

[24] Q. Mahmood, M. Yaseen, B. Ul Haq, A. Laref, and A. Nazir, Chem. Phys. **524**, 106 (2019).

[25] P. Giannozzi, S. Baroni, N. Bonini, M. Calandra, R. Car, C. Cavazzoni, D. Ceresoli, G. L. Chiarotti, M. Cococcioni, I. Dabo, et al., J. Condens. Matter Phys. **21**, 395502 (2009).

[26] J. P. Perdew, K. Burke, and M. Ernzerhof, Phys. Rev. Lett. **77**, 3865 (1996).

[27] P. E. Blöchl, Phys. Rev. B **50**, 17953 (1994).

[28] N. Holzwarth, A. Tackett, and G. Matthews, Comput. Phys. Commun. **135**, 329 (2001).

[29] R. M. Wentzcovitch, J. L. Martins, and G. D. Price, Phys. Rev. Lett. **70**, 3947 (1993).

[30] G. K. Madsen and D. J. Singh, Comput. Phys. Commun. **175**, 67 (2006).

[31] L. He, M. J. Tang, Y. Fang, and F. Q. Jing, Europhys. Lett. **83**, 39001 (2008).